# Angular dependence of magnetization reversal in epitaxial chromium telluride thin films with perpendicular magnetic anisotropy

Tanmoy Pramanik,* Anupam Roy,* Rik Dey,* Amritesh Rai, Samaresh Guchhait,† Hema CP Movva, Cheng-Chih Hsieh, and Sanjay K Banerjee

Microelectronics Research Center, The University of Texas at Austin, Austin, Texas 78758, USA

*Address correspondence to: pramanik.tanmoy@utexas.edu (T. P.), anupam@austin.utexas.edu (A. R.); T. P., A. R. and R. D. contributed equally to this work.
†Present addresses: Laboratory for Physical Sciences, College Park, Maryland 20740, USA and Department of Physics, University of Maryland, College Park, Maryland 20742, USA.

**ABSTRACT**

We investigate magnetic anisotropy and magnetization reversal mechanism in chromium telluride thin films grown by molecular beam epitaxy. We report existence of strong perpendicular anisotropy in these thin films, along with a relatively strong second order anisotropy contribution. The angular variation of the switching field observed from the magnetoresistance measurement is explained quantitatively using a one-dimensional defect model. The model reveals the relative roles of nucleation and pinning in the magnetization reversal, depending on the applied field orientation. Micromagnetic simulations are performed to visualize the domain structure and switching process.

## 1. INTRODUCTION

Increasing demand for data storage and data-intensive computing has triggered a tremendous research effort searching for materials and devices that hold the promise for ultra-high density, low-cost efficient storage [1,2] and non-volatile memory applications [3–5]. Hard disk drive (HDD), which stores information as the magnetization of the individual domains on the magnetic thin film media, has switched from longitudinal media (with in-plane remanent magnetization) to high anisotropy perpendicular media (with remanent out-of-plane magnetization) to increase the bit storage density. The retention time of the stored information in such perpendicular thin film media depends on the ratio of the anisotropy energy to the thermal energy, $K_u V/k_B T$, where $K_u$ is the perpendicular magnetic anisotropy (PMA) coefficient, $V$ is the volume of the bit, $k_B$ is the Boltzmann constant, and $T$ is the temperature [1]. Moreover, it has been shown that magnets possessing PMA can reduce the critical switching current in spin-transfer-torque random access memory (STT-RAM) to a lower value while meeting the long-term data retention time criterion over in-plane magnetic materials [6]. Materials with strong PMA are thus desired to increase the memory density in both cases. Recent studies of chromium telluride thin films have shown presence of high magnetic anisotropy in these materials [7,8]. In this paper, we study the magnetic anisotropy and magnetization reversal process in epitaxial chromium telluride thin films which possesses strong magneto-crystalline anisotropy.

Chromium telluride ($Cr_{1-\delta}Te$) material systems have been studied previously to investigate their suitability for different applications in spintronics [9,10] as well as for complex magnetic properties [11–15]. Different stable stoichiometry of these compounds have metal deficient NiAs-type crystal structure, depending on the value of $\delta$, and are ferromagnetic metals with a Curie temperature varying from 180 K to 340 K [16]. Early studies have focused on the crystal structure and magnetic moments of different phases in bulk form [11,13,16,17], while recent studies have focused more on thin films suitable for device applications, *e.g.*, intrinsic exchange bias [18], possibility of half-metallic ferromagnetism [7], and electric field modulation of ferromagnetism [10]. In our previous study, we reported presence of strong PMA in molecular beam epitaxy (MBE) grown $Cr_2Te_3$ thin films [8]. Here, we focus on the magnetic anisotropy and magnetization reversal process of epitaxial $Cr_2Te_3$ thin films. Interestingly, we find that the thin films show uniaxial anisotropy along the *c*-axis with a rather strong second order (second nonzero anisotropy constant) uniaxial anisotropy term, along with the first order (first nonzero anisotropy constant) term. The angular dependence of the switching field shows a complex behavior which can be satisfactorily explained by assuming a one-dimensional micromagnetic model of a defect and considering both inhomogeneous nucleation and domain wall pinning. We also investigate the magnetization switching dynamics and remanent domain structure in these thin films using micromagnetic simulations.

## 2. EXPERIMENTAL

$Cr_2Te_3$ films were grown in a custom-built MBE growth system (Omicron, Germany) under ultra-high vacuum (UHV) conditions (base pressure ~$1\times10^{-10}$ mbar). Details of the system and the growth method have been described elsewhere [8,19]. A reflection high energy electron diffraction (RHEED) setup is attached to the MBE system for *in situ* monitoring of surface reconstruction and growth. After pre-cleaning in acetone and isopropanol, P-doped n-type Si(111) wafers (oriented within ±0.5°) with a resistivity of 1-20 Ω-cm were introduced into the UHV chamber and atomically clean, reconstructed Si(111)-(7×7) surfaces were prepared by the usual heating and flashing procedure. Clean substrate surfaces were examined by *in situ* RHEED. Chromium and tellurium fluxes generated by *e*-beam evaporator and effusion cell, respectively, were co-deposited onto the substrates at an elevated substrate temperature of about 340 °C. The $Te_2$/Cr BEP (beam equivalent pressure) flux ratio was kept at about 15 and the chamber pressure during growth never exceeded $1\times10^{-9}$ mbar.

Post-growth investigations were performed by *in situ* RHEED operated at 13 kV, scanning tunneling microscopy (STM) at room temperature (RT) in the constant current mode, and X-ray photoelectron spectroscopy (XPS) with a monochromatic Al-K$\alpha$ source ($hv$ = 1486.7 eV). Magnetic and transport measurements were carried out with 9 T Quantum Design physical property measurement system (PPMS) combined with vibrating sample magnetometry (VSM) capable of cooling samples down to ~ 2 K.

## 3. RESULTS AND DISCUSSIONS
### 3.1 Growth and characterization

Figures 1 (a) and (b) show the RHEED patterns from a clean Si(111) surface reflecting the signature of (7×7) surface reconstruction for the electron beam along [1 1 -2]$_{Si}$ and [1 -1 0]$_{Si}$ incidence, respectively. Corresponding RHEED patterns from the same surface following growth of $Cr_2Te_3$ thin films are shown in Figures 1 (c) and (d). Sharp streaky features in RHEED patterns from the surface following growth of 4 nm $Cr_2Te_3$ thin film indicate high crystalline quality with atomically flat surface morphology. RHEED pattern also indicates the hexagonal structure of the layer grown along (001) direction (*c*-axis), which is also evident from the X-ray diffraction (XRD) pattern [8]. The growth structurally agrees very well with other reports of the growth of a thin film following hexagonal structure due to substrate crystal symmetry of a hcp(0001) and fcc(111) [19].

In Figure 1(e) and (f), we present an *in situ* STM study of the surface of $Cr_2Te_3$ thin film grown on Si(111)-(7×7) surfaces. The grown structures are characteristically triangular shaped because of the three fold symmetry of Si(111) surface. Careful investigation also reveals that the islands are triangular and/or truncated triangular, or, more precisely truncated hexagonal, with sharp edges or partially rounded in shape. The final shape of the structure depends on the variation in the growth rates along different edges of the structure. This also reflects the hexagonal crystal structure along the (001) direction, as both hcp(0001) and fcc(111) have the same hexagonal building blocks that differ only in the registry of the third layer. The morphologies and the structure agree very well with previous study [8]. Two STM images of scan areas 1500×1500 nm$^2$ and 1000×1000 nm$^2$ are shown in Figure 1(e) and (f), respectively, show several triangular and/or triangular hexagons and hexagonal surface lattices indicating the growth follows strictly the underlying crystal symmetry. STM images also show the presence of spirals and depressions in the film. Presence of threading dislocations with screw component leads to such spiral growth. Both the types of spirals, rotating clockwise and anticlockwise, are present in the film. One such spiral on a triangular hexagon structure is shown in the inset of Figure 1(f). The magnetic and magneto-transport measurements indicate the presence of short range magnetic ordering due to domain structure [8]. Morphology of the surface, as seen from STM studies, shows presence of features of different shapes and sizes which can lead to magnetic domain structures in the film. Presence of the geometrical imperfections, like the spirals due to threading dislocations, can also add up to the variation in magnetic homogeneity.

The elemental compositions of the grown films were investigated through *in situ* XPS. Survey spectrum of the film in Figure 2(a) shows presence of only Cr and Te as the constituent elements, which confirms the film to be free from the presence of any other elements as impurities. Figure 2(b) shows the high-resolution XPS spectrum of the Cr-2$p$ and Te-3$d$ peaks. Since the binding energies of these two peaks are very close to each other, it is difficult to separate them out into two components. Since the ratio of the photoionization cross-section of Cr-2$p$ and Te-3$d$ is about 2.67 for monochromatic Al-K$α$ X-ray source, Te-3$d$ peaks are dominant in the spectrum. The positions of the peaks are consistent with the reported results [20]. Figure 2(c) shows Cr-3$s$ core level spectra of $Cr_2Te_3$ thin film, which rests on the tail of a plasmon satellite accompanying the Cr-3$p$ peak. This agrees very well with the detailed photoemission study by Shimada *et. al.*[20]. Two broad peaks in the Cr-3$s$ spectrum indicates the itinerant nature of this compound and may originate from the exchange coupling between the 3$s$ core hole and 3$d$ band electrons. Several band structure calculations and other experiments have discussed the itinerant nature of chromium chalcogenide compounds [20]. Also, the separation between two peaks (~ 4 eV) is in agreement with that observed in different chromium chalcogenides [20].

### 3.2 Magnetic Properties

Magnetic and magneto-transport measurements carried out on $Cr_2Te_3$ thin films of different thicknesses have confirmed a ferromagnetic Curie temperature ($T_c$) of ~180 K and a strong magneto-crystalline PMA with the anisotropy axis along the out-of-plane $c$-axis [8]. Figure 3(a) and 3(b) show the in-plane and out-of-plane hysteresis loops, respectively, of a 20 nm thick film at 2 K. The square shape of the out-of-plane hysteresis loop with a coercive field much smaller than the in-plane saturation field indicates presence of strong PMA along the $c$-axis. To extract the uniaxial anisotropy coefficients from the measured in-plane (hard axis) hysteresis loop, we follow the method proposed by Sucksmith and Thompson [21]. Hysteresis in the in-plane magnetization curve near zero field presumably indicates formation of multi-domain state [22]. Deviation of the in-plane magnetization curve from that of coherent rotation due to formation of multi-domain state may affect the estimation of anisotropy coefficients by Sucksmith-Thompson method which assumes coherent rotation. However, as suggested in Ref. [22], anisotropy coefficients should be estimated from the fit to the non-hysteretic part of the magnetization curve. Thus, we have excluded the field range where the hysteresis occurs and fit only the data obtained at fields where there is no identifiable hysteresis (shown in the inset of Figure 3(a)). Data points at very high fields, which mark the onset of saturation have been excluded from the fitting as well. As the coherent rotation mechanism is valid in the selected range of magnetic field, the Sucksmith-Thomson method should yield a reliable estimation of the anisotropy coefficients. Considering a saturation magnetization ($M_s$) of $620\times10^3$ A/m [8], the first and second order anisotropy coefficients extracted from the fit are $K_{u1} = 8.71\times10^5$ J/m$^3$ and $K_{u2} = 2.85\times10^5$ J/m$^3$, respectively. This yields a ratio of $K_{u2}/K_{u1} = 0.33$ and a hard axis saturation field of $H_{sat} = 3.8$ T. The fitted in-plane magnetization curve, also shown in Figure 3(a), agrees well with the experimental data. Previously, a strong second order uniaxial anisotropy with $|K_{u2}/K_{u1}| \approx 1$ at 296 K was reported in CrTe [17]. Comparable values of cubic anisotropy and uniaxial anisotropy coefficients were also reported for CrTe with zinc-blende structure [7]. The ratio $K_{u2}/K_{u1}$ in our thin film is similar to the value for bulk cobalt ($K_{u2}/K_{u1} \approx 0.28$) [23] and Co/Pt thin films ($K_{u2}/K_{u1} \approx 0.3$-0.4) [24], but much higher than values observed in Ta/CoFeB/MgO material systems ($K_{u2}/K_{u1} < 0.1$) [25].

### 3.3 Angular dependence of the switching field

In Figure 4, the angular dependence of switching field from a 4 nm thick film is shown as a function of the angle ($\theta$) between the direction of the magnetic field and the anisotropy axis (c-axis). We have inferred the switching fields (black squares in Figure 4) from the magnetoresistance measurements reported in previous work (see Figure 5(b) in Ref. [8]). The magnetoresistance shows hysteresis with two sharp maxima at the same magnitude of positive and negative field values that correspond to the switching fields ($H_s$) [8]. As observed, the saturation field at $\theta = 90°$ is ~3.7 T, which is very close to the hard axis saturation field of 3.8 T for the 20 nm thick film (Figure 3). Assuming the same ratio of $K_{u2}/K_{u1} = 0.33$ and a saturation field of $H_{sat} = 3.7$ T, the anisotropy energies are calculated to be $K_{u1} = 8.39\times10^5$ J/m$^3$ and $K_{u2} = 2.75\times10^5$ J/m$^3$. From the angular dependence of the switching field, we first attempt to explain the magnetization reversal mechanism based on existing theoretical models. The switching field $H_s(\theta)$ in a Stoner-Wohlfarth (SW) coherent rotation switching mechanism [26] is given by $H_s(\theta) = H_{sat}[\sin^{2/3}(\theta)+\cos^{2/3}(\theta)]^{-3/2}$. The angular variation predicted by this equation shown in Figure 4(a) (curve C1) clearly indicates a different switching mechanism being responsible for the observed angular dependence. With a second order anisotropy term incorporated into the SW model [27], the resulting extended model is still insufficient [curve C2 in Figure 4(a)] and very different from the one observed in the experiment. This is not unexpected, as the coherent rotation models are only applicable for smaller particles with dimensions of the order of single domain particle. A more probable mechanism for magnetization reversal in extended thin films is by nucleation and propagation of reverse domain(s) [28,29]. In this case the switching field is decided by either of the two related mechanisms: nucleation of reverse domains or propagation of domain walls. As long as the reversal is limited by the nucleation field, the angular dependence may still follow a SW-like behavior even for larger particles [30]. For a thin film, the propagation of a domain wall can be hindered due to imperfections and thus the magnetization reversal process follows the Kondorsky model governed by the domain wall pinning strength [31]. The Kondorsky model has the well-known inverse cosine angular dependence given by $H_s(\theta) = H_s(0°)/\cos(\theta)$. As can be seen from Figure 4, the experimental data in our case does not follow the Kondorsky model [curve C3 in Figure 4(a)].

We observe that the switching fields are lower than the anisotropy field which is also well-known for other hard magnetic materials such as Sm-Co or Nd-Fe-B [32]. Such deviation can be explained considering nucleation and domain wall pinning due to defects and inhomogeneities in the material [32–35]. Following Sakuma *et. al.*[35], we assume a simple one-dimensional model with a defect of width $D$ in an otherwise perfect host region infinitely extended in either direction as shown in Fig 4(b) inset. We consider dimensionless parameters: $E = \frac{A^{II} K^{II}_{u1eff}}{A^I K^I_{u1eff}}$, $F = \frac{A^{II} M^{II}_s}{A^I M^I_s}$, $G = \frac{A^{II} K^{II}_{u2}}{A^I K^I_{u2}}$, where $A$ denotes the exchange constant, $K_{u1eff} = K_{u1} - \frac{1}{2} \mu_0 M_s^2$. Superscripts I and II denotes the host and the defect regions, respectively, and the defect region is considered to be magnetically weaker than the host ($0 \leq E, F \leq 1$). We also assume that both the first and second order anisotropy energies are lowered by the same ratio in the defect giving $E = G$ instead of $G = 0$ as has been considered by Sakuma *et. al.* [36]. From a uniformly magnetized initial state, as the external magnetic field is increased to reverse the magnetization, magnetization inside the defect region starts to follow the external field more closely than the host region. The maximum field above which a reverse domain appears inside the defect i.e., the magnetization inside the defect points towards the external magnetic field, is defined as the nucleation field. The pinning field is defined as the maximum field above which the defect region fails to limit the growth of an already reversed domain into the host regions. Depending on the values of $E$, $F$, $D$ and $\theta$, a reversed domain inside the defect could either escape into the host region, triggering magnetization reversal at the nucleation field or remain pinned inside the defect region until the value of the pinning field is reached, and the higher of these two fields decides the switching field. For each set of $E$, $F$ and $D$, the nucleation and pinning fields are obtained as a function of $\theta$ and compared with the experimentally observed angular dependence. The best fit obtained with $E = 0.23$, $F = 1.0$ and $D = 14$ nm, agrees well with the experimentally observed angular variation of the switching field [curve C4 in Figure 4(a)]. The corresponding nucleation and pinning fields are shown in Figure 4(b) which reveals that nucleation dominates the switching mechanism for $\theta < 70°$, above which the pinning field exceeds the nucleation field. This simple model adequately explains the comparative role of nucleation and pinning in the switching process.

### 3.4 Micromagnetic Simulation:

We attempt to picturize the magnetization reversal dynamics by means of micromagnetic simulations using MuMax3 micromagnetic simulator [37]. The micro-structure of the thin film is modeled in the simulation as randomly organized isosceles right triangles [Figure 5(a)]. Our assumption is valid since the STM morphologies of the film shows random azimuthal distribution of triangular or hexagonal features of sizes 50 nm – 100 nm and their merging in various ways to form a continuous film (Figure 1). We believe that this random orientation can produce a variation in the arrangement of the spins, giving rise to a randomly distributed domain structure with the spins inside each domain oriented in the same direction. Presence of spiral-like features can also add up to this cause. Existence of such imperfections suggests that a variation in the magnetic properties may also exist from island to island, as well as at the boundaries. The surface roughness estimated from STM imaging is found to be less than 1 nm and thus neglected in the micromagnetic model.

For the simulations, a total of 242 grains were considered which results into simulation geometry of 704 nm × 704 nm × 4 nm. A space discretization size of 1.375 nm is considered along $x$ and $y$ directions and a space discretization of 2 nm is considered along $z$ direction. Periodic boundary conditions are applied along the $x$ and the $y$ directions. The exchange energy constant used is 10 pJ/m, and both the saturation magnetization and exchange constant are assumed to be homogeneous throughout the thin film, as we have obtained $F = 1$ for the best fit considering the one-dimensional model. The anisotropy energies (both $K_{u1}$ and $K_{u2}$) are assumed to be the same within each grain, but vary from grain to grain following a Gaussian distribution with a variation of 5 % of the corresponding average values. For each grain the uniaxial anisotropy axis is also assumed to be randomly oriented with a tilt of 5° from the out of plane direction. A rectangular defect region of width $D = 14$ nm is considered at the boundary of two triangular grains [Figure 4(a)]. The material parameters inside the defect region are set such that $E = 0.23$ and $F = 1$ are fulfilled. An energy minimization solver [37,38] is used to simulate the hysteresis loops and a time integration solver with a damping constant of 0.5 has been used to simulate magnetization dynamics. We have considered $T = 0$ K for all the simulations.

Figure 5(b) shows the simulated hysteresis loops for the magnetic field applied at different angles from the out-of-plane direction. The in-plane hysteresis loop ($\theta = 90°$), shown in Figure 5(b), agrees with the experiment (Figure 3) qualitatively and shows similar narrow loop opening. The remanence state is always uniformly magnetized along ±z direction for $\theta \leq 80°$, whereas, multiple domain walls are observed in the in-plane ($\theta = 90°$) remanence state as shown in Figure 5(c) and 5(d). This also explains the higher magnetoresistance observed in the experiment at zero field for $\theta = 90°$ as compared to $\theta \leq 80°$ [8].

The magnetization configuration at different stages during switching for $\theta = 10°$ and $\theta = 75°$ are shown in Figure 6 (a)-(d) and (e)-(h), respectively. For both the cases, the nucleation is expected to initiate at a location where the anisotropy is much lower compared to the neighboring grains. In our simulation, the defect region has the lowest anisotropy energy [see Figure 5(a)], and thus acts as the nucleation center. Figure 6(a) shows the initial (time = 0 ns) magnetization configuration at the switching field. The magnetization inside the defect [inset in Figure 6(a)] starts deviating from the rest of the film and initiates the nucleation as time progresses [Figure 6(b)]. As the pinning field is lower than the nucleation field at $\theta = 10°$, the nucleated reverse domain grows out of the defect region and gradually propagates to complete the switching as seen from Figure 6 (b)-(d). At $\theta = 75°$, the initial magnetization configuration in Figure 6(e) shows an already nucleated domain [inset in Figure 6(e)] inside the defect. However, as the pinning field is higher than the nucleation field in this case, the reversed domain remained pinned in lower applied fields (not shown). When the magnetic field is increased further to the pinning field [shown in Figure 6(e)], the reverse domain overcomes the energy barrier and the domain wall gradually propagates to complete the magnetization reversal as seen from Figure 6 (f)-(h). The angular dependence of the switching field obtained from the micromagnetic simulations, shown in Figure 4(a) (curve C5), also agrees well with the experimentally observed values. We note that the specific morphological features and defects can also contribute to the switching behavior. However, our simple model with a 1D defect satisfactorily explains the angular dependence of the switching field observed in the experiment.

## 4. SUMMARY AND CONCLUSION

In conclusion, we have studied the magnetization reversal behavior of chromium telluride thin films grown by MBE. Though the *in situ* characterization using RHEED shows sharp streaky features indicative of smooth atomically flat surface film growth along the *c*-axis, STM measurements shows several crystalline imperfections which can lead to a domain structure of the grown film. Magnetic and magnetoresistance measurements confirm the film consists of domains. The angular dependence of the magnetization reversal in the film exhibits a complex nature. We show that a simple one-dimensional micromagnetic model of a defect can explain the angular variation satisfactorily. The magnetization switching dynamics studied using micromagnetic simulations confirmed the relative role of nucleation and pinning processes.

Strong perpendicular anisotropy in $Cr_2Te_3$ thin films could be useful for achieving thermally robust nanomagnets for spintronics applications as well as for high coercivity rare-earth-free material for permanent magnet applications. The present study hopefully sheds light on the dynamics of the magnetization reversal of a permanent magnet for possible applications in spintronics.

**Acknowledgement**
This was supported in part by the NRI SWAN center, NASCENT NSF ERC and the NSF NNCI program. The authors acknowledge the Texas Advanced Computing Center (TACC) at The University of Texas at Austin for providing computing facilities.

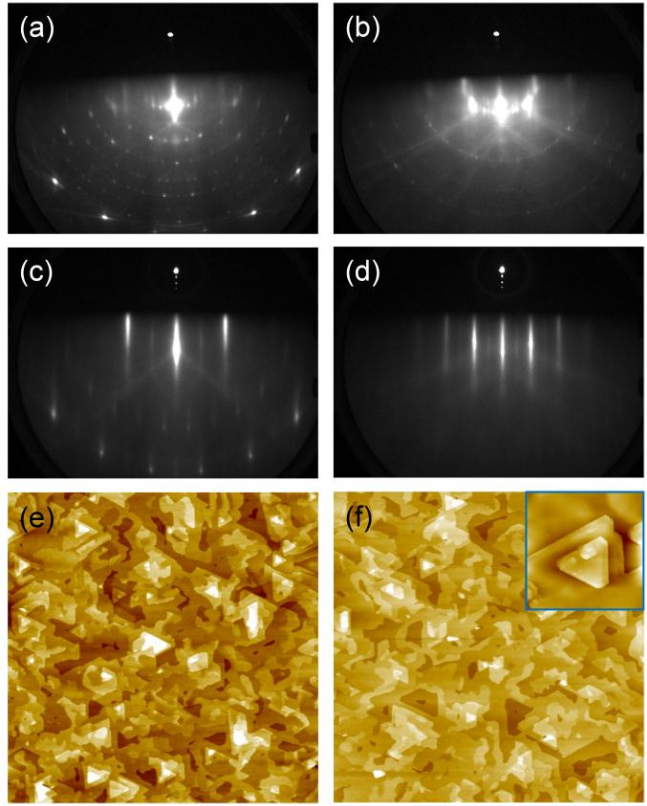

**Figure 1.** RHEED images following $Cr_2Te_3$ growth on Si(111)-(7×7) surfaces. (a) and (b) Typical (7×7) surface reconstruction from Si(111) substrate along [1 1 -2] and [1 -1 0] orientations of Si(111), respectively. (c) and (d) Corresponding RHEED patterns following 4 nm of $Cr_2Te_3$ growth. RT STM study from an epitaxial $Cr_2Te_3$ thin film at bias voltage: -0.8 V and tunneling current: 0.7 nA. Several triangular and truncated hexagonal features along with spirals are seen in (e) Scan area: 1500×1500 nm$^2$ and (f) Scan area: 1000×1000 nm$^2$. (inset) Spiral on a truncated hexagon (Scan area: 150×150 nm$^2$).

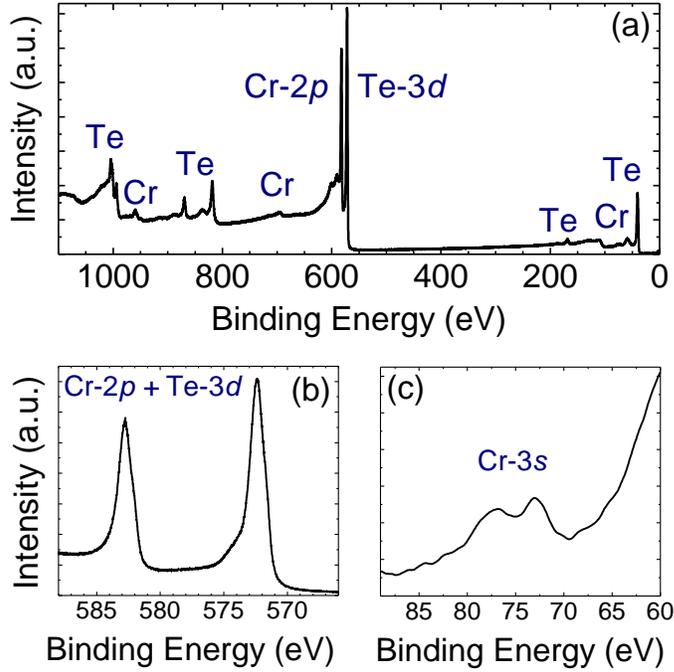

**Figure 2.** X-ray photoelectron spectroscopy (XPS) study from epitaxial $Cr_2Te_3$ thin film. (a) Survey spectrum showing only Cr and Te as constituent elements present in the film. (b) Cr-2*p* and Te-3*d* and (c) Cr-3*s* core-level x-ray photoelectron spectra of $Cr_2Te_3$ thin film.

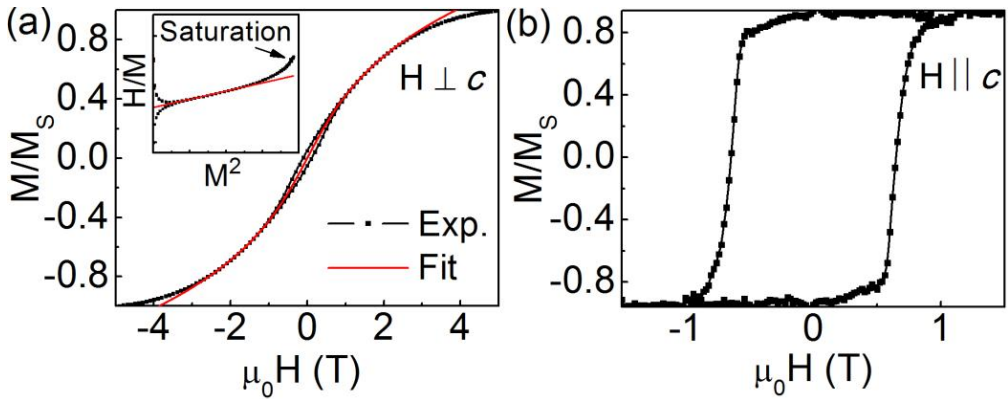

**Figure 3.** VSM measurements on a 20 nm thick film at 2 K. (a) Hysteresis loop (Black squares) measured with magnetic field applied in the in-plane (hard axis) direction. Analytical model (Red solid line) assuming coherent rotation with first and second order anisotropy terms agrees well with the magnetization curve. The values of the first and second order anisotropy terms are extracted from the y-intercept and the slope, respectively, of the straight line fit to the data as shown in the inset. (b) Hysteresis loop measured with magnetic field applied in the out-of-plane (easy axis) direction.

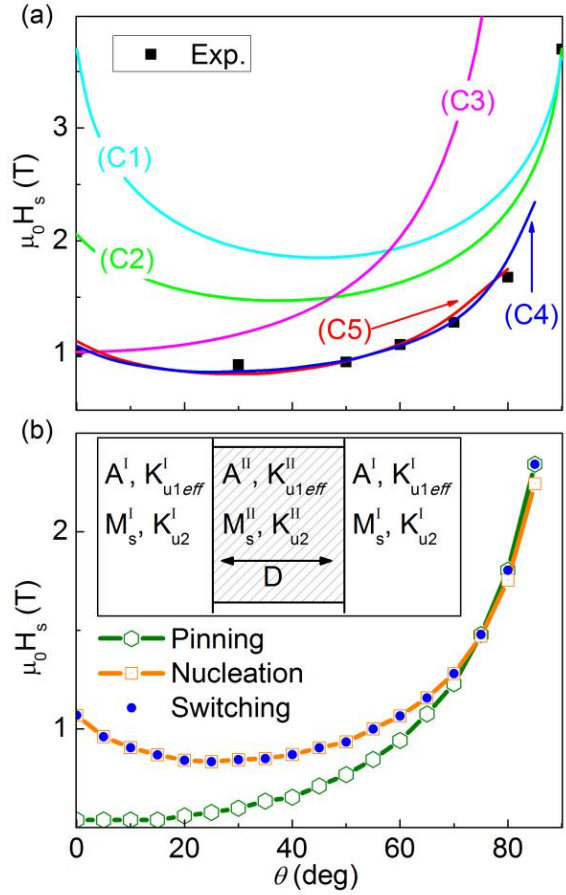

**Figure 4.** (a) Angular dependence of the switching field obtained from the magneto-transport measurement of a 4 nm thick film at 2 K. Experimental data points are shown by black solid squares. SW model considering only first order anisotropy (C1) and both first and second order anisotropies (C2). Kondorsky model showing a simple $1/\cos(\theta)$ dependence (C3). The experimental angular dependence is best explained by the one-dimensional defect model considering both nucleation and pinning processes (C4). Switching fields obtained from micromagnetic simulations (C5). (b) Angle dependence of the nucleation field (squares) and pinning field (hexagons) obtained from the one-dimensional model of a defect (inset) with the best fit values of the model parameters. The switching fields (circle) coincide with the nucleation fields up to $\theta = 70°$ and with the pinning fields for $\theta > 70°$.

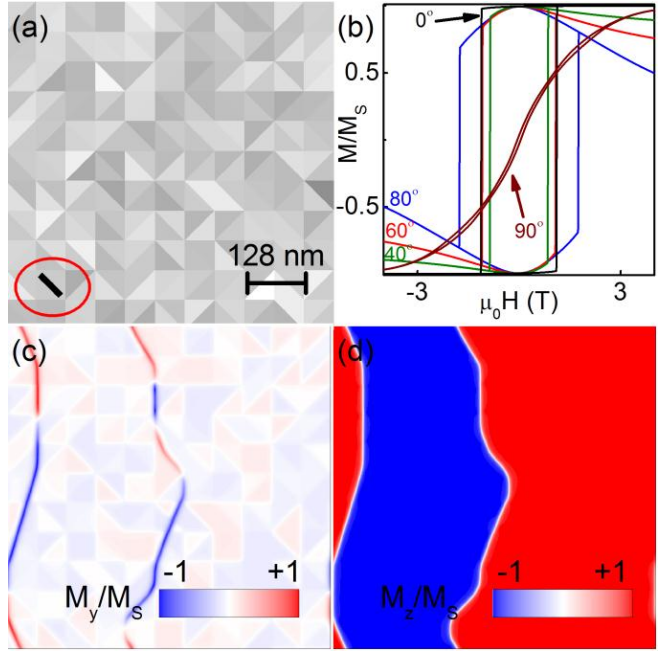

**Figure 5.** (a) The variation of the anisotropy energy $K_{u1}$ over randomly organized triangular grains considered in the micromagnetic model. Black to White corresponds to minimum to maximum value of the anisotropy energy. The grain with the defect is marked by the red circle. (b) Simulated hysteresis loops for different orientation of the magnetic field. Average out-of-plane component of magnetization ($M_z$) shows rectangular hysteresis loops for magnetic field orientations $\theta \leq 80°$. For $\theta = 90°$, average in-plane component of magnetization ($M_x$) is plotted. Domain structure formed at remanence after saturation in the applied magnetic field with $\theta = 90°$: (c) in-plane component of magnetization ($M_y$) shows domain walls, (d) out-of-plane component of magnetization ($M_z$) shows oppositely magnetized domains. The scale bar is same for (a), (c) and (d).

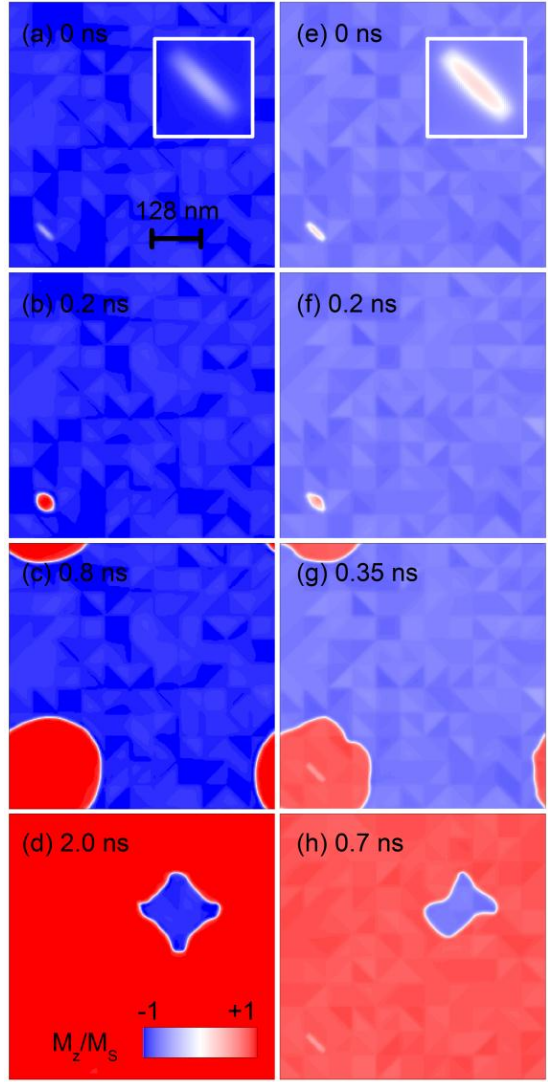

**Figure 6.** Magnetization dynamics observed in the micromagnetic simulation during switching from -z magnetized state to +z magnetized state. (a)-(d) show switching by reverse domain nucleation and domain wall propagation at an applied field of 0.93 T along $\theta = 10°$ direction. Inset of (a) shows the magnetization inside the defect region at the onset of nucleation. (e)-(h) shows the switching by domain wall depinning and propagation at an applied field of 1.56 T along $\theta = 75°$ direction. Inset of (e) shows the presence of an already nucleated reversed domain inside the defect. (a)-(h) follows the same color scale shown in (d). Length scale of (a)-(h) is shown in (a).